\documentclass[11pt]{article}

\usepackage[T1]{fontenc}
\usepackage[utf8]{inputenc}
\usepackage[english]{babel}

\usepackage{amsmath,amssymb}
\usepackage{braket}
\usepackage{physics}
\usepackage{siunitx}
\usepackage{subcaption}
\usepackage{graphicx}
\usepackage{graphicx}
\graphicspath{{Figures_RadicalSensing_arXiv/}}
\usepackage{float}

\usepackage[a4paper,margin=2.5cm]{geometry}
\usepackage{booktabs}  

\usepackage[colorlinks=true,
            linkcolor=blue,
            citecolor=blue,
            urlcolor=blue]{hyperref}

\usepackage[numbers]{natbib}

\title{Towards Application of Nanodiamonds for in-situ Monitoring of Radicals in Liquid Phase Chemical Reactions} 

\author{
Emma Herbst$^{1}$, Sebastian Westrich$^{1}$, Alena Erlenbach$^{1}$,\\
Jonas Gutsche$^{1}$, Maria Wächtler$^{2}$, Elke Neu$^{1}$\thanks{Corresponding author: nruffing@rptu.de}
}

\begin{document}

\maketitle 

\begin{center}
\small
$^{1}$Department of Physics and State Research Center OPTIMAS, RPTU University Kaiserslautern-Landau,
Erwin-Schroedinger-Str. 46, 67663 Kaiserslautern, Germany\\[0.3em]
$^{2}$Institute of Physical Chemistry, Kiel University, Max-Eyth-Str. 1, 24118 Kiel, Germany,\\
Kiel Nano, Surface and Interface Science (KiNSIS), Kiel University, Christian-Albrechts-Platz 4, 24118 Kiel, Germany
\end{center}

\begin{abstract}

In many chemical reactions, short-lived radical intermediates play a crucial role, while detecting such short-lived species in-situ remains challenging. The optically readable electronic spin of nitrogen-vacancy (NV) centers in diamond is a nanoscale sensor for such radical species: its longitudinal spin relaxation time  (T$_{1}$) reacts to magnetic fluctuations from the unpaired electrons of radical species in its local environment. In this setting, we demonstrate the successful in-situ detection of the nitroxide radical 2,2,6,6-Tetramethylpiperidinyloxyl (TEMPO) using NV center-based T$_1$ relaxometry after depositing nanodiamonds onto the inner wall of a glass cuvette. A significant concentration-dependent shortening of the relaxation time was observed, from 197\:\textmu s $\pm$ 21\:\textmu s without radical to 66\:\textmu s $\pm$ 30\:\textmu s at a concentration of 1\:M TEMPO. At the same time, the nanodiamonds remain firmly attached to the cuvette even after the solution under investigation has been exchanged. The detection is sensitive in the nanomolar (nM) range and the determined signal-to-noise ratio is between 1.6 and 3.

\end{abstract}

\section{Introduction}

Nitrogen Vacancy (NV) centers combine high sensitivity to magnetic fluctuations (T$_1$ relaxometry) with the ability to operate at room temperature (and above) as well as in a liquid environment \cite{rondin2014,perona2020}. In the life sciences, NV relaxometry has been established in recent years as a tool to sense radicals within cells and organisms \cite{perona2020,Fan2025,schirhagl2014,ZHANG2025,Niora2026}. Moreover, the fluorescence of NV centers has been shown to be stable even within very demanding chemical environments and has e.g.\ been used for optical thermometry in acid base neutralization reactions \cite{li2024}. Therefore, NV centers are particularly attractive sensors for chemical and catalytic research, where paramagnetic intermediates such as short-lived radicals or transition metal complexes play a central role \cite{radu2020}. The direct observation of such intermediates with unpaired electrons has so far been an experimental challenge, as they usually occur only transiently and their concentrations are extremely low. Traditionally, they are detected ex-situ using time-resolved spectroscopic methods such as electron paramagnetic resonance, stopped-flow, or online mass spectrometry in combination with microfluidics \cite{pahchan2025}. The superior magnetic sensitivity of NV centers opens up new possibilities here: changes in the local magnetic noise allow even a few paramagnetic species to be detected under realistic conditions while the chemical reaction of interest is running. This allows chemically relevant processes to be investigated in situ and under conditions close to those found in reality, which represents a decisive advantage over classical spectroscopic methods \cite{Lu2026}. The advantage arises from the combination of fluorescence-based measurements with the local sensitivity of the NV spin, as the relaxation time responds directly to magnetic fluctuations in the vicinity of the NV center.\\
The present work aims to demonstrate the feasibility of in-situ detection of free radicals in liquid phase using NV center-based T$_1$ relaxometry inside cuvettes. Cuvettes are a common approach for spectroscopy in chemistry, that allow simultaneously  for defined reaction conditions e.g.\ an oxygen free atmosphere as well as optical access for absorption and fluorescence measurements. Using individual nanodiamonds enables sub-micron resolution, which is crucial in spatially heterogeneous systems.
T$_1$ relaxometry, in particular, enables microwave free, low-back action detection of magnetic fluctuations in the range from megahertz to gigahertz, depending on the applied static magnetic bias field \cite{perona2020,hall2016,steinert2013,wood2016}. This microwave-free operation is especially advantageous for the cuvette-based approach employed in this work: Most microwave antenna assemblies used for NV electronic spin manipulation rely on close proximity of the antenna structure to the diamond used as a sensor. Typical layouts include $\Omega$-shapes antennas and striplines. In our case, the 1\,mm wall thickness of the cuvette inhibits efficient coupling of microwave radiation to the NV layer located on its inner surface via such typical antenna structures. We however do not exclude that our cuvette approach can be combined with local microwave delivers: Future studies could e.g.\ investigate the integration of an antenna patterned into the cuvette wall. Consequently, all optical T$_1$ relaxometry provides a promising strategy for the local in situ detection of reactive intermediates in chemical reactions, such as photocatalytic processes, and may contribute to a deeper understanding of their underlying reaction mechanisms \cite{Yang2026}.\\
While single crystal diamonds can provide a highly defined environment for NV centers, allowing optimal sensitivity (orientation, low strain etc.), we here intentionally investigate a nanodiamond approach for its higher applicability: cuvettes typically have at least millimeter long, thin necks through which a bulk diamond can be hardly inserted in a controlled way (e.g. using tweezers). Our work is also motivated by investigating an easy to apply, robust, cheap approach to radical sensing. From this point of view, we consider nanodiamonds advantageous here. \\
Typical applications for detecting radicals in liquids via NV centers have previously ranged from analyzing liquids in microfluidic devices on top of bulk diamonds with near-surface NV centers \cite{Marie2019, ziem2013, bucher2022} to drop-casting the liquid under investigation onto bulk diamonds \cite{steinert2013, Pham2011, Iyer2024}. Although applications using ensembles of nanodiamonds in cuvettes have been demonstrated recently \cite{Ralph2026}, no straightforward robust sensing approaches using individual nanodiamonds have been presented to measure radicals in solutions inside a cuvette yet. \\
Various paramagnetic species have to date been investigated using relaxometry: manganese ions \cite{ziem2013,mcguinness2013}, ferritin \cite{ziem2013,ermakova2013,schafernolte2014} and most prominently the magnetic resonance imaging (MRI) contrast agent gadolinium \cite{steinert2013,mamin2013,rendler2017,tetienne2013,sushkov2014}.\\
This work investigates the nitroxide radical 2,2,6,6-Tetramethylpiperidinyloxyl (TEMPO) as a model system which is also known as a spin-label in EPR. While we use TEMPO as an oxygen-stable radical for our proof-of-concept study here, our presented method can be modified to detect radicals under more challenging environments, e.g.\ oxygen free inert conditions, as often necessary in photocatalytic reactions.

\section{Experimental Section}

The following section describes the setup and the procedure used to perform T$_1$ relaxometry in liquid phase inside a cuvette. 

\subsection{Experimental setup}

To investigate the interaction between NV centers in commercially available nanodiamonds (NDNV-70nmHiFND, Adamas Technologies) and free nitroxide radicals (TEMPO) in an ethanol solution, a custom-built confocal scanning fluorescence microscope (compare \cite{pandey2022}) was employed. 
For fluorescence scanning and mapping, a diode laser at 520\:nm (DLnSec, Heilbronn Trading Company GmbH, 40\:mW maximum power) is utilized as excitation source at a power of 4\:mW. The nanodiamonds are positioned within the focal plane of a microscope objective [LUCPLFLN40X, Olympus] while we use lateral scanning of the sample to identify and investigate individual fluorescent spots. The investigated fluorescent spots show diffraction limited size (laterally and axially). While this does not fully exclude small agglomerates of nanodiamonds it provides, together with a low density of nanodiamonds, strong evidence that we observe individual nanodiamonds in our cuvette. We also note that DLS measurements performed by the supplier, indicate a very narrow size distribution of the nanodiamonds in the solution we start from. To avoid choosing agglomerates we only chose nanodiamonds within a certain magnitude of brightness ($\approx 10^{6}\,\mathrm{counts\:per\:second}$) and FWHM (700\,nm $\pm$ 100\,nm). Lateral scanning as well as placement in the focal plane was performed using a piezoelectric stage (ANSxyz100, Attocube Systems). The fluorescence emitted by the NV centers inside the nanodiamonds is collected via the same objective, ensuring that the excitation and emission radiation share the same optical path. The emitted signal is then routed through two identical dichroic mirrors and a 550 nm long-pass filter, to effectively remove backscattered excitation laser light from the beam path.
An avalanche photodiode (APD) [SPCM-AQRH-14-FC, Excelitas Technologies] is used to detect the NV fluorescence photons. For spectroscopic analysis, the detection path can alternatively be switched to a spectrometer [SP-2500i, Princeton Instruments]. In this case, the photons are dispersed according to their wavelength and detected by a CCD camera [Si-CCD sensor, PIXIS 256E Camera], to record the emission spectrum.

We here perform all-optical measurements of the NV center T$_{1}$ time: The NV center electronic spin is initialized into the $\mathrm{m_s=0}$ state using a first laser pulse. After a waiting time, the remaining population in $\mathrm{m_s=0}$  is monitored using a read-out pulse. While we are aware of the fact that changes in the charge state of NV centers can affect such all-optical T$_{1}$ measurements \cite{cardoso2023}, we here chose this approach as it can be straightforwardly implemented using the nanodiamonds in the cuvette, without the need to deliver microwaves for spin manipulation into the cuvette. The normalized readout signal exhibits a monoexponential decay, indicating that charge state changes can be negleted here, which can be described by \textbf{Equation \ref{eq:T1}}:

\begin{equation}
    \mathrm{N_{Relax}}(\tau)=C_a+C_be^{-\tau/T_1}.
    \label{eq:T1}
\end{equation}

Here, $\tau$ denotes the time between initialization and readout of the spin state, whilst the decay constant T$_{1}$ indicates the lifetime of the initialized spin state. The T$_{1}$ times of four different nanodiamonds for each concentration are recorded (compare \textbf{subsection \ref{subsec:T1}}) in order to obtain an average {T$_{1}$}.

\subsection{Sample preparation}

\begin{figure}[h!]
  \includegraphics[width=0.5\linewidth]{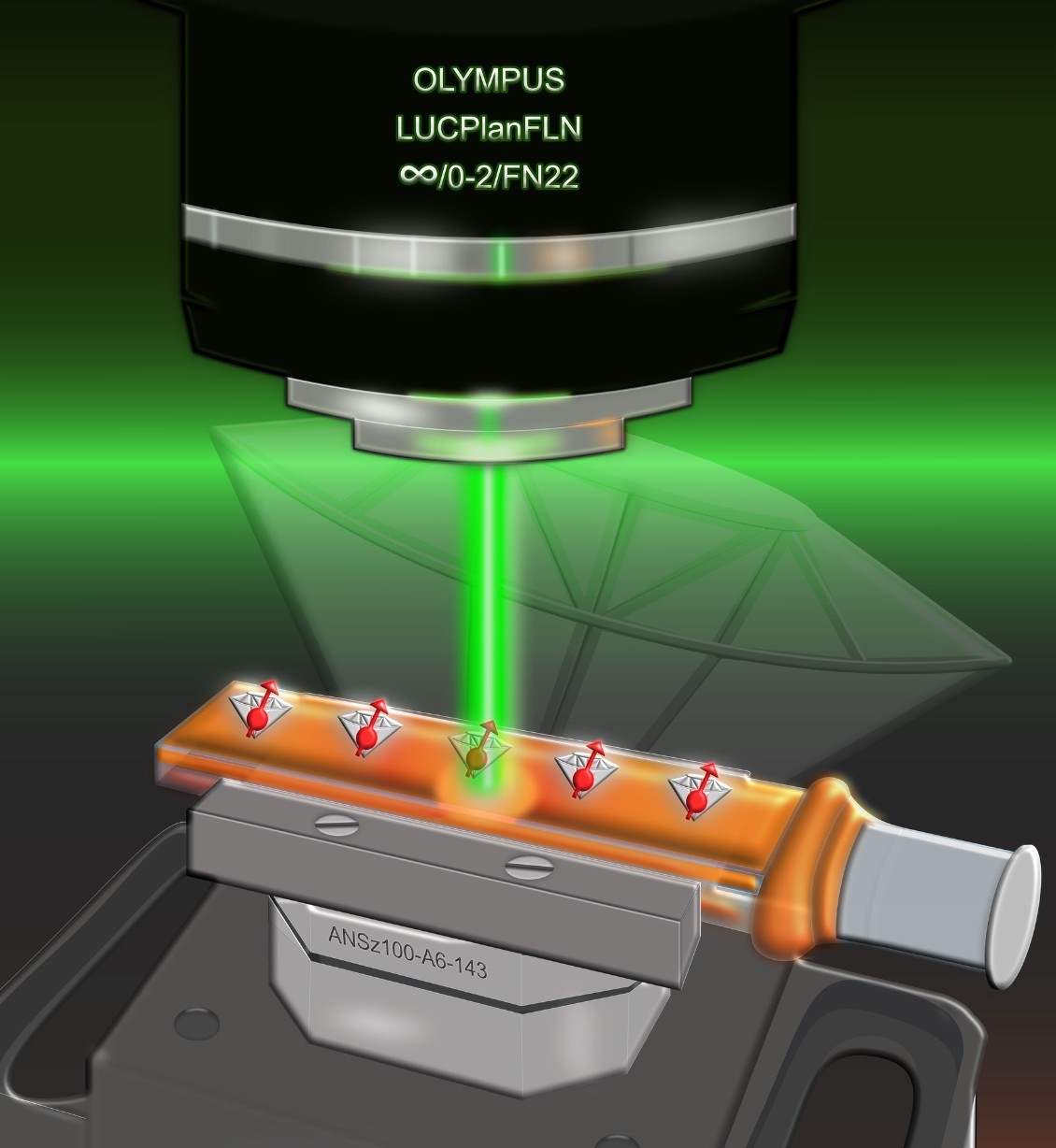}
  \caption{Illustration of the measurement setup: Nanodiamonds are spin-coated onto the inner upper wall of a cuvette, which is in the illustration filled with a high concentration of TEMPO diluted in ethanol (orange). After closing the lid, the cuvette is placed in a specifically designed sample holder so that it can be measured using the confocal microscope setup.}
  \label{fig:SETUP}
\end{figure}

To prepare the cuvette [Type 110-QS, HP quartz glass, wavelength range 200–2500\:nm, coating thickness 1\:mm, PTFE lid, 350\:\textmu L] for nanodiamond deposition, it is immersed for 5 minutes each in acetone, isopropanol and water. The cuvette is then dried on a hotplate at 120 $\mathrm{^\circ C}$ for 15\:minutes. This is followed by treatment in a plasma asher (100\:W O$_2$) [PICO-UHP, Diener electronic], after which the cuvette is placed back on the hotplate for another 5\:minutes.
Nanodiamonds suspended in water with a particle size of 70\:nm from Adamas Technologies are homogenized using a vortex mixer before being removed from the manufacturer’s container. Then a dilution is prepared for which 100\:\textmu L of the nanodiamond suspension are mixed with 900\:\textmu L of ultrapure water. Immediately before transferring the nanodiamonds into the cuvette, the nanodiamond dilution is homogenized again.
Three drops (approx. 200\:\textmu L) of the nanodiamond suspension are pipetted onto one of the inner walls of the cuvette. We place the cuvette on a spin coater [Delta 80BM Gyrset, Süss MicroTec] and rotate it at 3000\:rpm for 20\:minutes. During the spin coating process, the majority of water evaporates, yet to ensure complete removal of residual water, the cuvette is placed on a hotplate until no further condensation is evident.
Afterwards, 400\:\textmu L of TEMPO diluted in ethanol as solvent can be filled directly into the cuvette, which is then closed with a lid and wrapped with parafilm to ensure it is tightly sealed. A concentration series of TEMPO diluted in ethanol is prepared (compare \textbf{Table \ref{tab:TEMPO}}) and then measured, starting with the lowest (nM) concentration and proceeding to the highest (M) concentration. We point out that we did not re-apply the nanodiamonds when exchanging the TEMPO solution. We also did not observe a change in the density of the nanodiamonds after we exchanged the solution. This is a very strong indication that the nanodiamonds' attachment to the cuvette is very robust making our approach practical to reliably perform repeated sensing of radicals in cuvettes functionalized using nanodiamonds. An illustrative image of the cuvette filled with a high concentration of TEMPO under the confocal microscope is shown in \textbf{Figure \ref{fig:SETUP}}.

\begin{table}[h!]
 \caption{Preparation of the concentration series for the radical TEMPO (M = 156.25\:$\frac{\mathrm{g}}{\mathrm{mol}}$).}
 \label{tab:TEMPO}
 \begin{tabular}{@{}llll@{}}
  \hline
  Step & Starting solution & Volumes & Concentration \\
  \hline
  High conc. 
    & Solid material (165,34\:mg) 
    & 1,00\:mL 
    & $\approx 1\,\mathrm{mol\,L^{-1}}$ \\
  
  Stock solution 
    & Solid material (0,16\:mg) 
    & 1,00\:mL 
    & $\approx 1 \cdot 10^{-3}\,\mathrm{mol\,L^{-1}}$ \\

  1st Dilution 
    & 1\:mM Solution 
    & 2\,\textmu L + 998\,\textmu L Ethanol 
    & $\approx 2  \cdot 10^{-6}\,\mathrm{mol\,L^{-1}}$ \\

  2nd Dilution 
    & 2\:\textmu M Solution 
    & 2\,\textmu L + 998\,\textmu L Ethanol 
    & $\approx 4 \cdot 10^{-9}\,\mathrm{mol\,L^{-1}}$ \\

  3rd Dilution 
    & 4\:nM Solution 
    & 250\,\textmu L + 750\,\textmu L Ethanol 
    & $\approx 1 \cdot 10^{-9}\,\mathrm{mol\,L^{-1}}$ \\
    
  \hline
 \end{tabular}
\end{table}

\section{Results and Discussion}

This section summarizes our findings regarding fluorescence spectra, confocal fluorescence maps, T$_1$ relaxometry, and calculated signal-to-noise ratios of the introduced samples.

\subsection{Fluorescence spectra of NDNV70nmHiFND and TEMPO} 
\label{subsec:Relaxometry}

As highly concentrated TEMPO solutions show an orange-red color, we first check whether the solution emits fluorescence light under 520\:nm laser excitation in the same spectral region as the NV center fluorescence, as this could interfere with the spin read-out of the NV centers. Therefore, fluorescence spectra of several concentrations (micro- and nano- molar as well as molar concentration) of the TEMPO solution in the cuvette without nanodiamonds are recorded. 

\begin{figure}[h]
    \centering
    \begin{subfigure}[t]{0.49\textwidth}
        \centering
        \includegraphics[width=\linewidth]{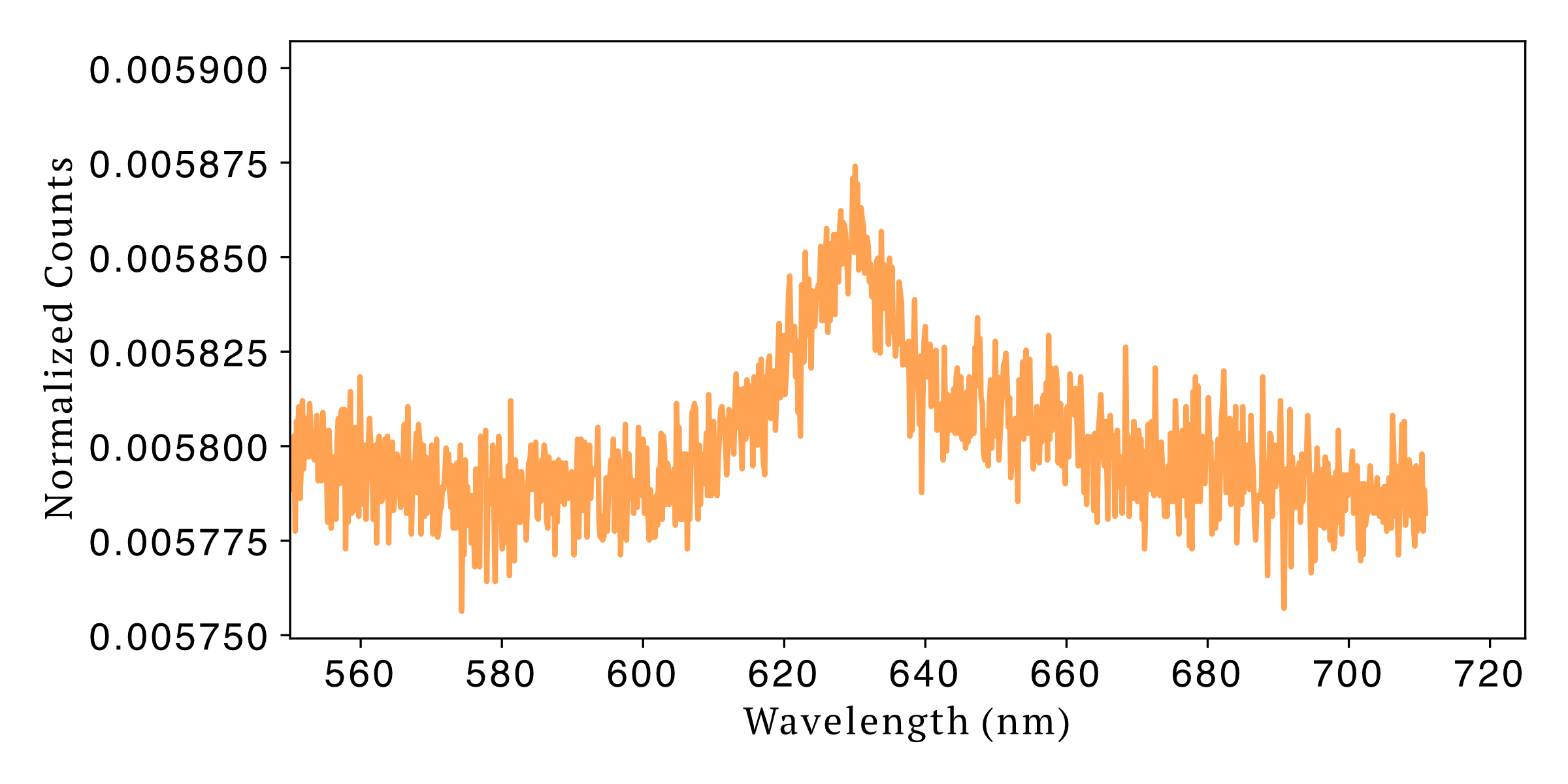}
        \caption{Fluorescence spectrum of the highest concentration of TEMPO (1\:M). The spectrum shows a peak at $\approx$ 630\:nm.}
        \label{fig:TEMPOFL}
    \end{subfigure}
    \hfill
    \begin{subfigure}[t]{0.49\textwidth}
        \centering
        \includegraphics[width=\linewidth]{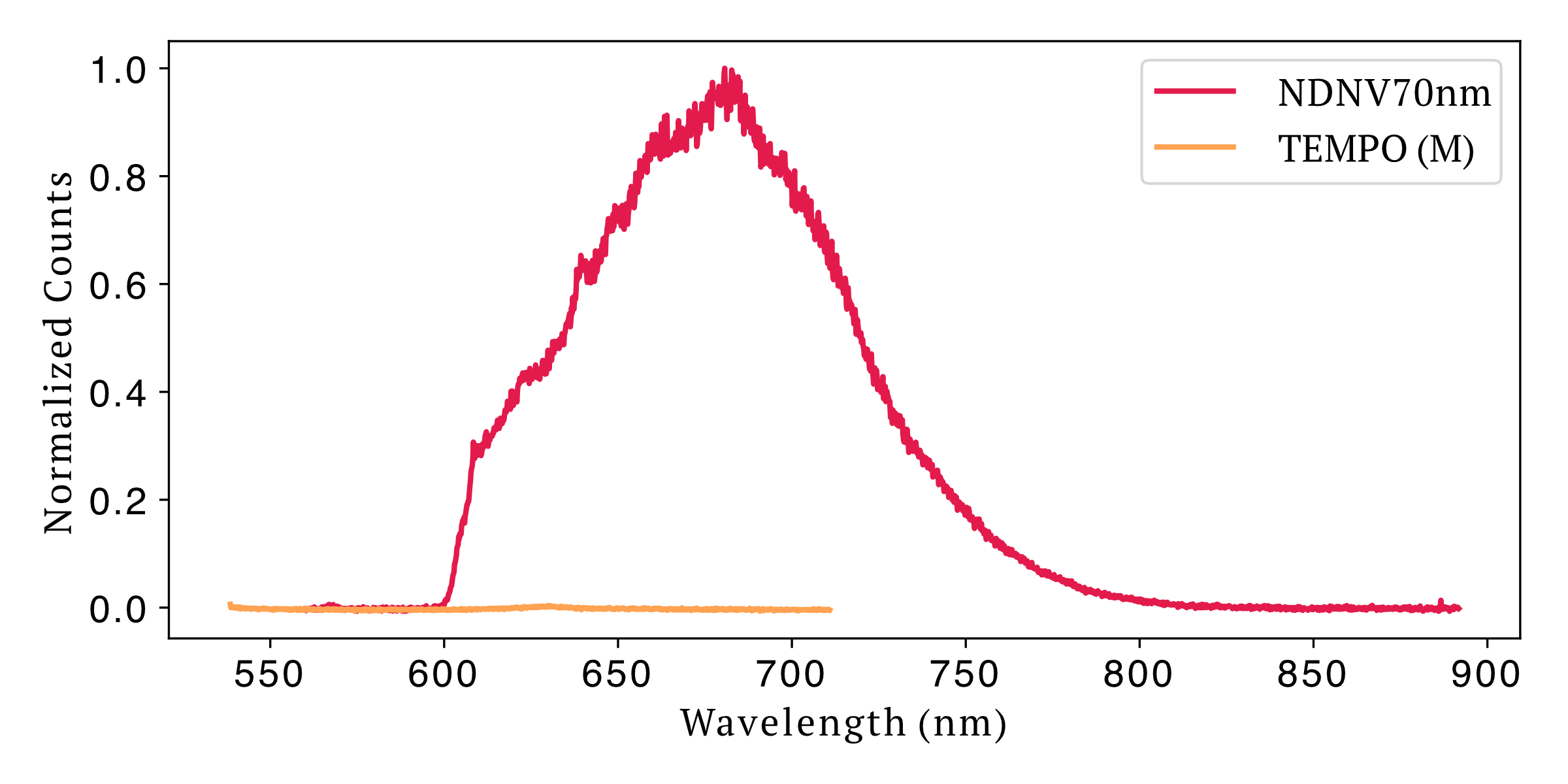}
        \caption{Normalized fluorescence spectra of TEMPO (orange) and NDNV70nm (red).}
        \label{fig:NVFL}
    \end{subfigure}
    \caption{Normalized fluorescence spectra of the different sample components.}
\end{figure}

\textbf{Figure \ref{fig:TEMPOFL}} exemplarily shows the fluorescence spectrum of the molar concentration TEMPO solution as we observed the highest emission for this concentration. The fluorescence spectra for the other concentrations are therefore not shown here; nevertheless, the data can be found in the dataset. While we observe an emission peak around 630\:nm in \textbf{Figure \ref{fig:TEMPOFL}}, the emitted light's intensity is so small ($\Delta \mathrm{Normalized\:Counts}_{\:630\:\mathrm{nm}} \approx 0,0001$) that it should not hinder the detection of the NV centers at different TEMPO concentrations (compare \textbf{Figure \ref{fig:NVFL}}). If the spectra are normalized not only to the area but also to the maximum of the NV spectrum, the TEMPO spectrum in orange appears as merely a line (see \textbf{Figure \ref{fig:NVFL}}). This makes it even clearer that TEMPO fluorescence should not hinder the detection of the NV centers. \\
The spectrum of the nanodiamonds with NV centers shown in red in \textbf{Figure \ref{fig:NVFL}} is in accordance with theory outlined by Rondin et al. (2014) \cite{rondin2014}, exhibiting a peak at $\approx$ 640\:nm and a broad region extending from 650\:nm to 725\:nm. This region corresponds to the zero-phonon line (ZPL) and the phonon sidebands of an NV$^{-}$ center. The ZPL of the NV$^{0}$ charge state can not be observed here, as we used a 550\,nm LP and 600\,nm cut off dichroics so that the signal under 600\,nm is cut off. We here chose this set of optics as we were interested to collect a large fraction of NV$^{-}$ fluorescence and not in spectroscopic investigation of the NVs. Consequently, due to the limited spectral range, we cannot infer a  NV$^{-}$/ NV$^{0}$ ratio.

\subsection{Confocal scans of the cuvette with high and low radical concentrations}

\begin{figure}[h!]
    \centering
    \begin{subfigure}[t]{0.49\textwidth}
        \centering
        \includegraphics[width=\linewidth]{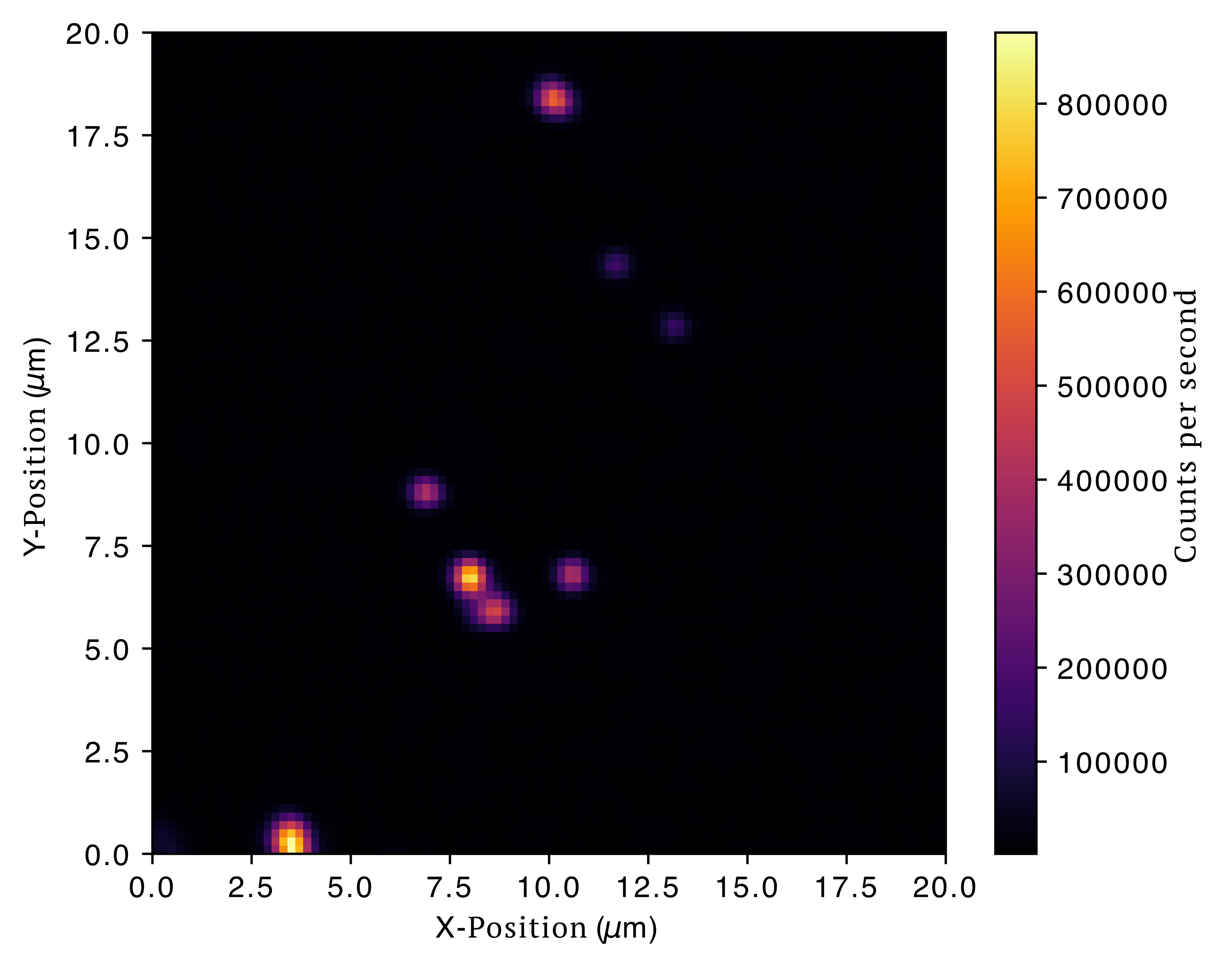}
        \caption{Confocal fluorescence map of nanodiamonds in the cuvette. Eight nanodiamonds can be seen within an area of 20 $\times$ 20\:\textmu m.}
        \label{fig:SCANLOW}
    \end{subfigure}
    \hfill
    \begin{subfigure}[t]{0.49\textwidth}
        \centering
        \includegraphics[width=\linewidth]{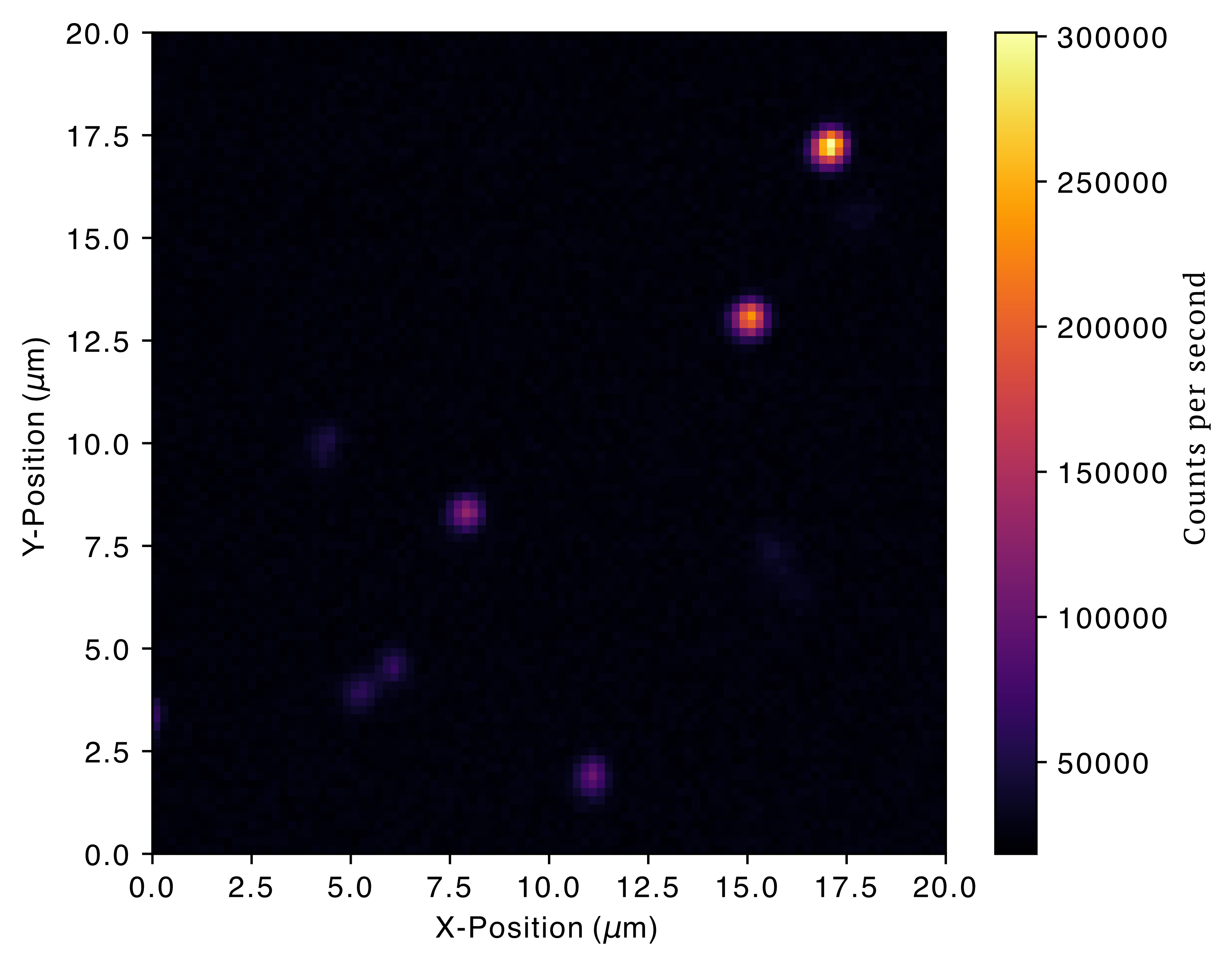}
        \caption{Confocal fluorescence map of of nanodiamonds with a molar radical concentration in the cuvette. Seven nanodiamonds can be seen within an area of 20 $\times$ 20\:\textmu m.}
        \label{fig:SCANHIGH}
    \end{subfigure}
    \caption{Confocal scans of samples using the same cuvette}.
\end{figure}

A confocal fluorescence map of nanodiamonds in the cuvette is depicted in \textbf{Figure \ref{fig:SCANLOW}}.
The examined area (20 $\times$ 20\:\textmu m) for this concentration contains eight nanodiamonds, which we subsequently investigate. We note that when the TEMPO solution is changed, the cuvette has to be removed from the holder in our confocal microscope. Because of the large size  (length 8\,cm and width 1\,cm of the cuvette), together with the manual re-mounting of the cuvette into our confocal microscope after we change the TEMPO solution, we were not able to identify the same nanodiamonds again. The patterns are not very characteristic, the nanodiamond density is low and the area as well as the uncertainty in re-mounting is large. Thus, the region investigated in the confocal fluorescence maps will not be identical for the different TEMPO concentrations and we can consequently not re-identify the same nanodiamonds. We however are sure that the nanodiamonds adhere to the cuvette as we always find them in the same depth corresponding to the inner wall of the cuvette and the density as well as the typical fluorescence signals do not change. The nanodiamonds are not reapplied during the course of the experiments; the same cuvette and the original nanodiamond coating are used throughout. Accordingly, the measurements presented in \textbf{Figure \ref{fig:SCANLOW}} and \textbf{Figure \ref{fig:SCANHIGH}} were performed using the same cuvette and nanodiamonds applied in the same coating procedures.

While our preliminary fluorescence investigations of the molar TEMPO solutions (compare \textbf{subsection \ref{subsec:Relaxometry}}) suggested that the fluorescence from the TEMPO solution should not infer with the observation of NV centers, we notice that it is harder to find NV nanodiamonds when the cuvette is filled with the molar concentration of TEMPO solution. We thus scan larger areas of the cuvette and identify two equally bright nanodiamonds in the magnitude of $\approx 10^{6}\,\mathrm{counts\:per\:second}$ (visible in \textbf{Figure \ref{fig:SCANHIGH}}). We tentatively assume that the loss in brightness we observe here is caused by the intense magnetic noise in the molar TEMPO solution.   

\subsection{Relaxometry of NDNV70nmHiFND in a  TEMPO solution} \label{subsec:T1}

\begin{figure}[h]
    \includegraphics[width=0.7\textwidth]{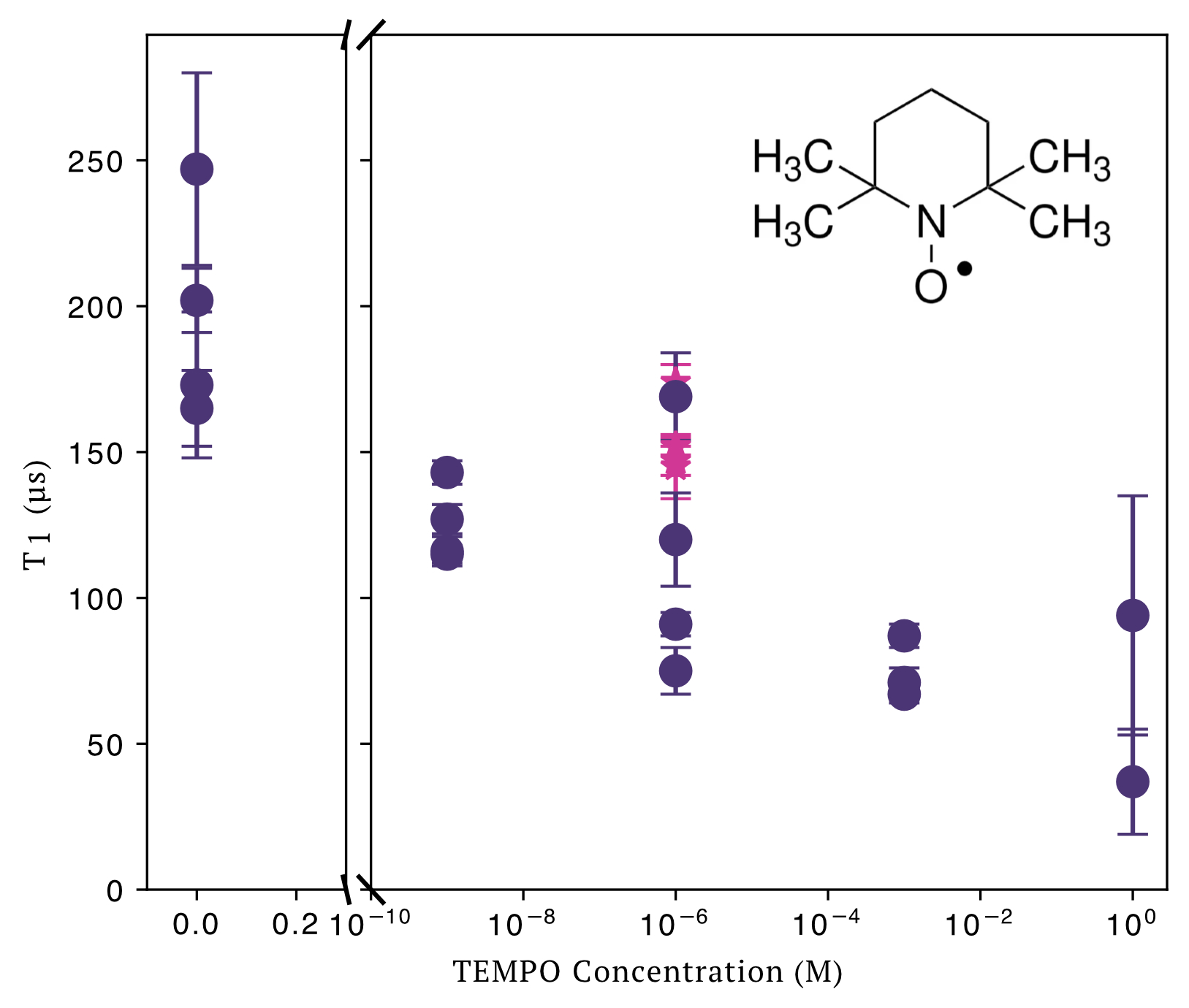}
    \caption{Results of all-optical T$_{1}$ measurements using freshly prepared TEMPO solutions (shown in purple dots) and a solution six days old (shown in pink stars). The figure shows the T$_{1}$ time as a function of TEMPO concentration. A higher concentration results in a decrease of the T$_{1}$ time. The inset in the upper right shows the structural formula of 2,2,6,6-Tetramethylpiperidinyloxyl (TEMPO) \cite{tempo}.}
	\label{fig:T1Kuv}
\end{figure}

First, the all-optical measurement of the T$_{1}$ relaxation time of nanodiamonds in the cuvette without a radical is carried out, with a total of four nanodiamonds being measured. This is followed by the all-optical measurement of the T$_{1}$ relaxation time in ascending concentrations of the TEMPO radical dissolved in ethanol, with four nanodiamonds measured for each concentration. Details of the preparation can be found in \textbf{Table \ref{tab:TEMPO}}.\\
The resulting T$_{1}$ times in \textmu s are shown in \textbf{Figure \ref{fig:T1Kuv}} as a function of the TEMPO concentration in M. Measurements using freshly prepared solutions are shown in purple, those using a six-day-old solution in pink. For the molar concentration and the $10^{-3}$ molar concentration, we found overall three nanodiamonds with T$_{1}$ values higher than the average value we found without the TEMPO solution. We have not plotted these three datapoints but nevertheless list them in \textbf{Table \ref{tab:T1}} in red together with their error ($\Delta$ T$_{1}$). \\
We tentatively attribute this observation to the fact that especially for the molar concentration, we had to search for the brightest nanodiamonds that might also correspond to the largest nanodiamonds. Here, NV centers on average have a larger distance to the nanodiamond surface, rendering them less sensitive to magnetic fluctuations in the surrounding solution but also less sensitive to fluctuations at the nanodiamond surfaces themselves thus leading to higher T$_{1}$. The inherent spread of T$_{1}$ times for individual nanodiamonds constitutes a disadvantage in our approach but does not hinder to observe a clear trend. For future experiments, markers could be fabricated that aid in re-identifying individual nanodiamonds. For future applications, we also plan to trigger radical generation by illumination, while the cuvette stays mounted in the setup.\\

The observed decrease in the mean relaxation time from 197\:\textmu s $\pm$ 21\:\textmu s without TEMPO to 66\:\textmu s $\pm$ 30\:\textmu s at a TEMPO concentration of 1 M is of a comparable order of magnitude to that observed in similar experiments with Gd$^{3+}$ \cite{Iyer2024}.

\begin{table}[h!]
 \caption{T$_{1}$ time in dependency of the TEMPO concentration.}
 \label{tab:T1}
 \begin{tabular}{@{}lcc@{}}
  \hline
  TEMPO concentration (M) & T$_{1}$ (\textmu s) & $\Delta$ T$_{1}$ (\textmu s) \\
  \hline
  0 & 165, 173, 202, 247 & 13, 25, 11, 33 \\
  $1 \cdot 10^{-9}$ & 115, 116, 127, 143 & 3, 5, 5, 4 \\
  $1 \cdot 10^{-6}$ & 152, 147, 174, 145 & 3, 5, 6, 11 \\
  $2 \cdot 10^{-6}$ & 169, 120, 75, 91 & 15, 16, 8, 4 \\
  $1 \cdot 10^{-3}$ & 67, 87, 71, \textcolor{red}{252} & 3, 4, 5, \textcolor{red}{156} \\
  $1 \cdot 10^{0}$ & 37, 94, \textcolor{red}{506}, \textcolor{red}{834} & 18, 41, \textcolor{red}{140}, \textcolor{red}{170} \\
  \hline
 \end{tabular}
\end{table}

\subsection{Signal-to-Noise Ratio}

Analogous to Steinert et al. (2013) \cite{steinert2013}, a signal-to-noise ratio (SNR) can be calculated for the relaxometry measurements evaluated in \textbf{Subsection \ref{subsec:Relaxometry}}. This is achieved by dividing the mean T$_1$ time of the nanodiamonds in the cuvette by the mean T$_1$ time of the respective radical concentration. The resulting values are shown in \textbf{Table \ref{tab:SNR}} and are in the range of $\mathrm{SNR} = 1.6-3$. For comparison, Steinert et al. in \cite{steinert2013} achieved a significantly higher SNR ratio of 15.5 at a concentration of 1 M Gd$^{3+}$.\\
The difference essentially stems from the fact that Steinert et al. (2013) \cite{steinert2013} employed a wide-field setup with single crystal CVD diamonds hosting shallow NV centers at a mean depth of 7\:nm.
Compared to nanodiamonds, single crystal CVD diamonds have a significantly higher crystal quality and a lower surface defect density, resulting in longer T$_1$ times.
The initial T$_1$ time is approximately six times larger than the value measured in \textbf{Subsection \ref{subsec:Relaxometry}}.

\begin{table}[h!]
 \caption{Relationship between radical concentration, T$_1$ time and SNR in NV centers in nanodiamonds.}
 \label{tab:SNR}
 \begin{tabular}{@{}ccc@{}}
  \hline
  Concentration & T$_1$ (\textmu s) & SNR \\
  \hline
  0     & $197 \pm 36$ & -- \\
  nM    & $125 \pm 13$ & 1.58 \\
  \textmu M & $114 \pm 41$ & 1.73 \\
  mM    & $75 \pm 11$  & 2.63 \\
  M     & $66 \pm 40$  & 2.98 \\
  \hline
 \end{tabular}
\end{table}

\section{Conclusion}

The present study investigated the nitroxide radical TEMPO as a model system to assess the feasibility of robustly applying nanodiamonds in vessels typically used for observing chemical reactions spectroscopically. We thus proof the feasibility to detect radicals in-situ in such systems using quantum sensing.\\
To this end, the nanodiamonds were fixed to the inner surfaces of a cuvette using spin coating, so that nanodiamonds could be repeatedly observed using a confocal fluorescence microscope. The respective sample was then added as a liquid and examined in increasing concentrations with regard to the T$_1$ relaxation time of the NV centers. The attachment of the nanodiamonds was so robust that it allowed for multiple exchanges of the liquid inside the cuvette without a discernible reduction of nanodiamond density.\\
In our experiment, the all-optical in-situ T$_1$ relaxometry shows a clear concentration-dependent shortening of the relaxation time, from 197\:\textmu s without the radical to 66\:\textmu s at a concentration of 1 M TEMPO despite the fact that we could not track the same nanodiamonds for different concentrations. We thus demonstrate a tool capable of detecting radicals in liquids in the nanomolar range.\\ 
Our approach to apply nanodiamond inside a cuvette can also be applied to chemical reactions sensitive to oxygen, which require a protected environment. This is particularly relevant given the geometric constraints of the cuvette system, which limit the direct introduction of bulk diamond samples, especially for long-neck inert cuvettes required for oxygen-free measurements. In contrast, nanodiamonds can be introduced in a flexible and non-invasive manner, enabling their use even under such demanding experimental conditions. Consequently, this approach provides a versatile route towards in situ quantum sensing in a broad range of chemically relevant, oxygen-sensitive systems.

\section*{Supporting Information} 
All Data used in this manuscript is publicly available via the Zenodo repository under the following DOI: doi.org/10.5281/zenodo.19594502.

\section*{Author Contributions} 
E.H. conceived and conducted the study, performed the investigation and data curation, developed the methodology, created the visualizations, and wrote the original draft of the manuscript. S.W. and A.E. contributed to the investigation and methodology and provided resources. J.G., E.N. and M.W. supervised the project, contributed to project administration and funding acquisition. All authors reviewed and edited the manuscript and approved the final version.

\section*{Conflict of interest} 
The authors declare no conflict of interest.

\section*{Acknowledgements}
The authors thank K. Rediger and S. Burai for their support with regard to the handling of the radical sample. Funding for this work was provided by the Deutsche Forschungsgemeinschaft (DFG, German Research Foundation) under Grant No. TRR 173–268565370, Spin+X (Project A12). E.H. acknowledges the GEQS project funded by the Carl-Zeiss-Stiftung. We acknowledge the Quantum Initiative Rhineland-Palatinate (QUIP). 

\newpage

\bibliographystyle{unsrt}
\bibliography{References}

\end{document}